\documentclass[runningheads,a4paper]{llncs}
\usepackage{mystyle}

\usepackage{url}
\urldef{\mailsa}\path|mateusz.lacki@biol.uw.edu.pl|
\urldef{\mailsb}\path|aniag@mimuw.edu.pl|

\newcommand{\keywords}[1]{\par\addvspace\baselineskip
\noindent\keywordname\enspace\ignorespaces#1}

\begin{document}
\mainmatter  

\title{Law of Localised Fine Structure}

\subtitle{\textit{with application in mass spectrometry}\thanks{
	This research was partially supported by Polish National Science Center grant $\text{n}^\text{o}$ 2011/01/B/NZ2/00864.
}}


\titlerunning{Mass Spec Fine Structure Distribution}

\author{Mateusz Krzysztof \L\k{a}cki
\and Anna Gambin}

\authorrunning{Fine Structure Distribution in Mass Spectrometry}

\institute{
  Faculty of Mathematics, Informatics and Mechanics\\ University of Warsaw, Banacha 2, 02-097 Warszawa, Poland\\
  \mailsa\\
  \mailsb
}


\toctitle{Mass Spec Fine Structure Distribution}
\tocauthor{Mateusz \L\k{a}cki}
\maketitle

\begin{abstract}
	This paper presents a brand new methodology to deal with isotopic fine structure calculations. By using the Poisson approximation in an entirely novel way, we introduce mathematical elegance into the discussion on the trade-off between resolution and tractability. Our considerations unify the concepts of fine-structure, equatransneutronic configurations, and aggregate isotopic structure in a natural and simple way. We show how to boost the theoretical resolution in a seemingly costless way by several orders of magnitude with respect to the already very efficient algorithms operating on isotopic aggregates. We also develop an effective new way to obtain the important peaks in the most disaggregated isotopic structure localised in a precise region in the mass domain.
\end{abstract}


\section{Introduction}

Recent advances in mass spectrometric technology allow for a more and more elaborate application in biology. It is being recognised that more precise information can be retrieved even from larger chemical compounds. More resolved spectra already now help in the identification of complex mixtures of biomolecules, such as proteins and peptides, nucleic acids, and drugs; see \cite{Milandanovic2012OnTheUtilityOfIsotopicFineStructure}. 

It is well known that part of their complexity stems from the existence of stable isotopes. It is because of them that a given analyte is represented as a series of peaks, rather than just one corresponding to its monoisotopic mass. Depending on the machine, the isotopic structure can be resolved at different levels of accuracy. This provides a rationale for development of efficient algorithms that calculate their theoretical counterparts.

In this paper we consider three basic levels of aggregation of the isotopic structure,  corresponding to three distinct levels of theoretical resolution: the most coarse clumps together peaks with the same additional nucleon count, cf. \cite{Kienitz1961MassSpectrometry}, the finer one distinguishes between the {\it equatransneutronic groupings}, see \cite{Olson2009Calculations}, while the finest one represents completely resolved isotopic configurations, see \cite{Rockwood1995Ultrahighspeed}. The theoretical underpinnings of how to mathematically model the impact of isotopes are already well established, see \cite{Valkenborg2012Isotopic}, and the probability of a given exact fine configuration can be obtained using the product of multinomial distributions. However, with the growth of molecule one observes a general rise in the complexity in the problem of enumerating all fine isotopic configurations. To bypass this problem, different simplifications were proposed, amounting to different ways of binning configurations together explicitly \cite{Claesen2012Efficient} or by hiding them under the guise of Fourier Transform \cite{Rockwood1995Relationship}. 

Here we propose two refinements over the aggregate model, as used in \cite{Claesen2012Efficient}. Both of them use the concept of the {\it localised fine structure}, which corresponds to isotopic configurations clustered together into only one peak under the aggregate model. One of the devised algorithms extremely efficiently disaggregates that peak into {\it equatransneutronic groupings}; the other one fully resolves the isotopic pattern. Both of these algorithms are based on elegant Poisson approximations to the generally acknowledged multinomial model. To our best knowledge this type of approximation have not yet been used for algorithmic purposes. It has been used however in the context of proteomic and peptide research: in \cite{Breen2000AutomaticPeak} it was used for high throughput protein identification and then it was reevaluated in \cite{Valkenborg2007UsingPoisson} for peptides. 
\section{Approximations}

By an isotopic configuration we understand information on numbers of different isotopes a chemical compound in the sample is made of. For the purpose of simplicity, we focus here on chemical compounds composed of carbon, hydrogen, nitrogen, oxygen, and sulfur; still, results of this section generalize to any compound whatsoever. Thus, we concentrate on compounds like \molecule, where the low case letters describe the numbers of atoms of particular element type. Among such compounds one can already find peptides and proteins. An isotopic configuration could be represented by an extended empirical formula, 
\begin{equation}\label{long chemical formula}
	\text{\moleculeIsotopic}.
\end{equation}

In the above representation, small letters with indices represent counts of different atoms with indices displaying the number of additional neutrons an isotope has with respect to the lightest possible isotopic variant. 

Rather than \eqref{long chemical formula}, we shall be using an equivalent probabilistic notation, treating upper case letters, like \ce{^{12}C}, as random variables and considering small case letters, $\cem{c_0}$, to be their realizations. An expression like $A = \{ \ce{^{13}C} = \cem{c_1},\, \ce{^{2}H} = \cem{h_1} \}$ is shorthand for saying: let us focus on all configurations \eqref{long chemical formula} that have \ce{c_1} heavy carbons and \ce{h_1} deuters in total.

Following \cite{Kienitz1961MassSpectrometry}, one assumes that the law of vector
\begin{equation}\label{long chemical vector}
	\left( \cem{^{12}C},\, \cem{^{13}C},\, \cem{^{1}H},\, \cem{^{2}H},\, \cem{^{14}N},\, \cem{^{15}N},\, \cem{^{16}O},\, \cem{^{17}O},\, \cem{^{18}O},\, \cem{^{32}S},\, \cem{^{33}S},\, \cem{^{34}S},\, \cem{^{36}S} \right),	
\end{equation}
given \molecule, is a product of independent multinomial distributions,
{\small\begin{equation}\label{product of multinomials}
	\MM = \mathrm{Multi} \Big( \prob(\cem{^{12}C}), \prob(\cem{^{13}C}); c \Big)
	\otimes \dots \otimes 
	\mathrm{Multi} \Big( \prob(\cem{^{32}S}), \prob(\cem{^{33}S}), \prob(\cem{^{34}S}), \prob(\cem{^{36}S}); s \Big),	
\end{equation}}
where the probabilities of observing particular isotopes, $\prob(\cem{^{12}C})$, \dots, $\prob(\cem{^{36}S})$, are established in independent experiments, cf. Table \ref{basic info on isotopes table}. For instance, the probability of a given carbons configuration $(\cem{c_0}, \cem{c_1})$ equals
$$
	\mathrm{Multi} \left( \prob(\cem{^{12}C}), \prob(\cem{^{13}C}); c \right)
		\Big( (\cem{c_0}, \cem{c_1}) \Big) = 
	\begin{pmatrix}
		\cem{c} \cr \cem{c_0}, \cem{c_1}  
	\end{pmatrix} \prob(\cem{^{12}C})^\cem{c_0} \prob(\cem{^{13}C})^\cem{c_1}
$$
and it should be multiplied by similar expression for hydrogen, nitrogen, oxygen and sulfur to obtain probability for configuration \eqref{long chemical formula}.

Observe, that given \molecule, part of the information in representation \eqref{long chemical vector} is redundant and can be shortened by neglecting counts of the lightest isotope variants, leaving us with 
\begin{equation}\label{short chemical vector}
 	\left( \cem{^{13}C},\, \cem{^{2}H},\, \cem{^{15}N},\, \cem{^{17}O},\, \cem{^{18}O},\, \cem{^{33}S},\, \cem{^{34}S},\, \cem{^{36}S} \right).	
\end{equation}
Missing terms can be retrieved from relationships $\cem{^{12}C} + \cem{^{13}C} = \cem{c}$, $\cem{^{1}H} + \cem{^{2}H} = \cem{h}$, and so on, that occur with probability one.

\begin{mydef}\label{localised fine structure definition}
	We call the set of configurations  
	{\small
		\begin{equation}\label{LFS_K}
			LFS_K	=
			\left\{ 
				\cem{^{13}C} + \cem{^{2}H} +  \cem{^{15}N} +  \cem{^{17}O} +  \cem{2 $\times$^{18}O} +  \cem{^{33}S} +  \cem{2 $\times$^{34}S} + \cem{4 $\times$^{36}S} = K	
			\right\}
		\end{equation}
	}
	a \emph{\textbf{localised fine structure} with $K$ extra neutrons}.  	
\end{mydef}

The reason for numbers 2 and 4 appearing above is that \ce{^{18}O} and \ce{^{34}S} have two additional neutrons, and \ce{^{36}S} -- four; cf. Table \ref{basic info on isotopes table}.

The problem of enumerating all elements of $LFS_K$ is known as the money exchange problem. In general, it corresponds to finding all integer solutions $(x_1, \dots, x_k)$ of a {\it Linear Diophantine Equation}  
\begin{equation}\label{Linear Diophantine Equation}
	d_1 x_1 + \dots + d_k x_k = K,
\end{equation}
where $(d_1, \dots, d_k)$ are integer coefficients. According to \cite{Agnarsson2002OnTheSylvesterDenumerants}, if the greatest common divisor of $(d_1, \dots, d_k)$ is equal to one, then the number of solutions to Eq. \eqref{Linear Diophantine Equation} is approximately $\frac{K^{k-1}}{(k-1)! d_1 \dots d_k}$. Carbon has only one additional isotope, so $\exists_i d_i = 1$ in \eqref{Linear Diophantine Equation}. The above estimate encompasses therefore all of organic chemistry and proteomics. 

Nonetheless, since configurations in $LFS_K$ are naturally prioritized by probability \eqref{product of multinomials} one would be satisfied with enumerating only the most probable ones. 


\begin{Problem}\label{Problem of finding LFS_K configurations.}
	For a given $K$, find a small set $B \subset LFS_K$ of configurations s.t. 
	\begin{equation}\label{problem equation}
		\MK (B) := \frac{ \MM(B) }{ \MM( LFS_K ) } \approx 1\,,	
	\end{equation} 
	where $\MK$ is the product of multinomial laws \eqref{product of multinomials} conditioned on the set of configurations in $LFS_K$ and is referred to as \emph{\textbf{The Law of Localised Fine Structure}}.
\end{Problem}

In statistical terms, we are interested in approximating some critical set of large probability, as measured by the {\it Law of Localised Fine Structure}.

Why should one study law described by \eqref{problem equation} in the first place? Simply because the masses of different configurations in $LFS_K$ concentrate around the compound's monoisotopic mass shifted to the right by $K$ Da; c.f \cite{Hughey2001KendrickMassDefect}. For medium sized compounds, $LFS_K$'s for different $K$ should in principle form disjoint clusters in the mass to charge domain, with some interference for bigger compounds. Studying $LFS_K$ guarantees exploration of a precise region in the mass to charge domain.

To solve Problem \ref{Problem of finding LFS_K configurations.} we approximate measure $\MK$ by a more analytically tractable measure $\QK$ defined on the $LFS_K$. We then devise an algorithm to find a possibly small set of configurations $B^* \subset LFS_K$, s.t. $\QK (B^*) \approx 1$. Since $\QK \approx \MK$, so $\MK (B^*) \approx 1$ and $B^*$ solves Problem \ref{Problem of finding LFS_K configurations.}, possibly suboptimally.

A natural way to define proper $\QK$ is to first approximate $\MM$ by some $\QQ$ and then pose $\QK (\circ) := \frac{\QQ (\circ \cap LFS_K) }{\QQ(LFS_K)}$, i.e. condition $\QQ$ on the occurrence of configurations from $LFS_K$. To prove it works, we have to first mention, that by approximation we understand convergence in distribution, as described in \cite{Kallenberg2002FoundationsOfModernProbability}. Then, we make use of the following lemma: 

\begin{lemma}\label{conditional convergence lemma}
	Let $\mu^{[n]}, \mu$ be discrete measures. If $\mu^{[n]}$ converges in distribution to $\mu$, $\mu^{[n]} \rightharpoonup  \mu$, and an event $A$ has nonzero probability under any of that measures, $\underset{n}{\forall} \mu^{[n]}(A)\,,\, \mu(A) > 0$, then measures conditioned by $A$, $\mu^{[n]}_A (\circ) := \frac{\mu^{[n]} ( \circ \cap A)}{\mu^{[n]}(A)}$ converge in distribution to $\mu_A (\circ) := \frac{ \mu( \circ \cap A) }{ \mu(A) }$; or $\mu^{[n]}_A \rightharpoonup \mu_A$ for short.
\end{lemma}  
Proof is to be found in \textbf{Appendix}.

Let us now unveil the usefulness of Lemma \ref{conditional convergence lemma}. There is an entire family of measures mentioned in it, $\mu^{[n]}$. We assume, that one of them is simply our initial measure: there exists $n^*$ s.t. $\MM = \mu^{[n^*]}$. Also, we assume the approximation of $\mu^{[n^*]}$ by measure $\mu$ is already o good one. Our choice for $\mu$ is to be the product of independent Poisson measures, which is stimulated by the following, well known lemma.

\begin{lemma}\label{weak convergence of multinomial to Poissons lemma}
	If all\,\,$\lim_{n\to \infty} n p_{k,n}= \lambda_k$ exist for $k \in \{1,\dots, w\}$, then 
	
	\begin{equation}\label{weak convergence of multionial to Poissons equation}
		\mathrm{Multi}\left( p_1^{[n]}, \dots, p_w^{[n]}; n \right) 
			\rightharpoonup 
		\mathrm{Poiss}( \lambda_1) \otimes \dots \otimes \mathrm{Poiss}( \lambda_w ),	
	\end{equation}
	where $\mathrm{Poiss}$ stands for the Poisson distribution, $\mathrm{Poiss}(\lambda)(k) 	= \frac{\lambda^k}{k!}e^{-\lambda}$.
	
\end{lemma}

In Lemma \ref{weak convergence of multinomial to Poissons lemma} one assumes that the number of trials $n$ goes to infinity. In our model this corresponds to an infinite enlargement of the compound. The existence of limits assumes that this enlargement is done so that on such an idealized compound only the lightest isotopes would appear infinitely often. Moreover, since the support of any Poisson distribution is equal to the set of all integer numbers, the state space of configurations gets significantly enlarged and contains configurations that are nonphysical for any real chemical compound. For instance, positive probabilities would be prescribed to configurations with numbers of isotopes greater then the number of possible places for them on any finite compound. Observe also, that the probabilities $p_k^{[n]}$ are pending towards zero: for good approximation one would expect therefore the probabilities of observing heavier isotopes, e.g. quantities like $\prob(\cem{^{13}C}), \prob(\cem{^{2}H}), \dots, \prob(\cem{^{36}S})$, to be relatively small. That is the case -- cf. Table \ref{basic info on isotopes table}.

Observe, that Lemma \ref{weak convergence of multinomial to Poissons lemma} defines a proper limit for just one multinomial distribution, whereas $\MM$ is a product thereof. The problem is other than what to do with products: one can approximate independently each multinomial. However, the quality of such approximation depends on all the counts of different elements in a molecule. For instance, in case of \molecule\,the better the approximation\footnote{The {\it goodness} of approximation is expressed in the total variance distance; see \cite{Roos1999OnTheRateOfMultivariatePoissonConvergence}.} the bigger the smallest among numbers $(\cem{c}, \cem{h}, \cem{n}, \cem{o}, \cem{s})$. Due to the polymer structure, one would expect some more information could be revealed on that matter for proteins and peptides. Indeed, empirical research by Senko et al. \cite{Senko1995Determination} established the concept of $m$-avergine, i.e. an averaged protein: any protein composed of $m$ amino acids should have its mass approximately equal to the mass of the idealised compound 
\begin{equation*}
	\cem{C}_{\lfloor m \times 4.9384\rfloor} 
	\cem{H}_{\lfloor m \times 7.7583\rfloor} 
	\cem{O}_{\lfloor m \times 1.4773\rfloor} 	
	\cem{N}_{\lfloor m \times 1.3577\rfloor} 
	\cem{S}_{\lfloor m \times 0.0417\rfloor}.
\end{equation*}

The weakest link in the approximation might result from small numbers of sulfur. This is an acknowledged problem in empirical studies, as exposed in \cite{Valkenborg2007UsingPoisson}. The longer the polymers however, the smaller the differences should be.

The final questions is: what values should be used as $\lambda$'s in Lemma \ref{weak convergence of multinomial to Poissons lemma}? We {\it calibrate} those values by equating them to the averages of the multinomial distributions from \eqref{product of multinomials}: in case of carbon we set $\lambda_\cem{^{13}C} \approx \cem{c} \times \prob( \cem{^{13}C} )$.  In contrast to our method, $\lambda$'s in \cite{Breen2000AutomaticPeak,Valkenborg2007UsingPoisson} are chosen to be the minimizers in a free parameter optimisation scheme with $\chi^2$ penalty\footnote{Note however, that these two solutions should not differ too much for larger compounds, for it is known that both the Poisson and Multinomial distributions are concentrated near their means, see \cite{Bobkov1998OnModifiedLogarithmicSobolev}.}.

All in all, the probability assigned to event
\begin{equation*}
 	\left\{ \cem{^{13}C} = \cem{c_1},\, \cem{^{2}H} = \cem{h_1},\, \cem{^{15}N} = \cem{n_1},\, \cem{^{17}O} = \cem{o_1},\, \cem{^{18}O} = \cem{o_2},\, \cem{^{33}S} = \cem{s_1},\, \cem{^{34}S} = \cem{s_2},\, \cem{^{36}S}= \cem{s_4} \right\}	
\end{equation*} 
is given by
\begin{equation}\label{QK Nominator}
	\poiss{\text{c}}{13}{1}
	\poiss{\text{h}}{2}{1}
	\poiss{\text{n}}{15}{1}
	\poiss{\text{o}}{17}{1}
	\poiss{\text{s}}{33}{1}
		e^{ - \mu}
	\poiss{\text{o}}{18}{2}	
	\poiss{\text{s}}{34}{2}
		e^{ - \eta }		
	\poiss{\text{s}}{36}{1}
		e^{ - \gamma },
\end{equation}
where 
\begin{align*}\label{intensities summed equation}
	\mu 	&=	\lambda_\cem{^{13}C} + \lambda_\cem{^{2}H} + \lambda_\cem{^{15}N} + \lambda_\cem{^{17}O} +\lambda_\cem{^{33}S}  	\\
	\eta 	&= 	\lambda_\cem{^{18}O} + \lambda_\cem{^{34}S}\\ 
	\gamma	&= 	\lambda_\cem{^{36}S}.
\end{align*}

The usefulness of approximation by a product of independent Poisson lies in two important properties, as summarised in the following lemmas.

\begin{lemma}\label{sum of independent Poissons lemma}
	Suppose we have a collection of $m$ independent Poisson-distributed random variables, $X_i \sim \mathrm{Poiss}(\kappa_i)$. Then $X_1 + \dots + X_m \sim \mathrm{Poiss}(\kappa_1 + \dots + \kappa_m)$. 
\end{lemma}  

\begin{lemma}\label{Poisson conditional on sum of Poissons}
	Suppose we have a collection of $m$ independent Poisson-distributed random variables, $X_i \sim \mathrm{Poiss}(\kappa_i)$. Then $X_1, \dots, X_m$ given that $X_1 + \dots + X_m = K$ is multinomially distributed,

$$ 
	\Big(X_1, \dots, X_m | X_1 + \dots + X_m = K \Big) 
	\sim 
	\mathrm{Multi}\Big( \frac{\kappa_1}{\sigma}, \dots, \frac{\kappa_m}{\sigma}; K \Big), 
$$
	where $\sigma = \sum_{i = 1}^m \kappa_i$.	
\end{lemma}
Both lemmas are proved in \cite{Kingman1993PoissonProcesses}. Lemma \ref{sum of independent Poissons lemma} shows how to simplify calculations for a Diophantine equations with all parameters set to one. Lemma \ref{Poisson conditional on sum of Poissons} describes the law resulting from conditioning independent Poisson variables by such an expression. 

Suppose that we concentrated on molecules composed entirely of elements that can have only one additional neutron, e.g. \smallMolecule. By Lemma \ref{Poisson conditional on sum of Poissons} we get:

\begin{result}\label{Multinomial Result}
 	For \smallMolecule, let $\tilde{\mu} := \lambda_\cem{^{13}C} + \lambda_\cem{^{2}H} + \lambda_\cem{^{15}N}$. Then
 	$$\QK = \mathrm{Multi}\left(
 		\frac{\lambda_\cem{^{13}C}}{\tilde{\mu}}, 
 		\frac{\lambda_\cem{^{2}H}}{\tilde{\mu}}, 
 		\frac{\lambda_\cem{^{15}N}}{\tilde{\mu}}; K \right).$$
\end{result}

\begin{proof}
	The corresponding Diophantine equation is $\cem{^{13}C} + \cem{^2H} + \cem{^{15}N} = K$.
\end{proof}

It is valuable to see, how Lemma \ref{Poisson conditional on sum of Poissons} generalizes while conditioning on a more complex Diophantine equation. Observe, that we can rewrite the definition of $LFS_K$ emphasizing the {\it equatransneutronic grouping}, i.e. gluing together counts of configurations with the same numbers of extra neutrons,
\begin{equation*}
	LFS_K = \Big\{ \underbrace{\cem{^{13}C} + \cem{^2H} + \cem{^{15}N} + \cem{^{17}O} + \cem{^{33}S}}_{ G_1 } + \,2 \times \underbrace{( \cem{^{18}O} + \cem{^{34}S} )}_{ G_2 } + \,4 \times \underbrace{\cem{^{36}S}}_{ G_4 } = K \Big\},	
\end{equation*}
so that in light of Lemma \ref{sum of independent Poissons lemma}, $\QQ(A)$ can be calculated in an easier way: 
$$\QQ( LFS_K ) = \sum_{k_1 + 2 k_2 + 4 k_4 = K} \prob( G_1 = k_1, G_2 = k_2, G_4 = k_4 ),$$
where $G_1 \sim \mathrm{Poiss}( \mu )$, $G_2 \sim \mathrm{Poiss}( \eta )$, and $G_4 \sim \mathrm{Poiss}( \gamma )$ are mutually independent. In light of \cite{Olson2009Calculations}, $G_i$ is equal to the total number of atoms bearing exactly $i$ additional neutrons.  

To calculate $\QK$ it remains to divide expression \eqref{QK Nominator} by $\QQ( LFS_K )$. Subsequent multiplication of both the nominator and the denominator of that result by $\frac{\mu^{k_1}}{k_1 !} \frac{\eta^{k_2}}{k_2 !} \frac{\gamma^{k_4}}{k_4 !}$ gives us an even more clear image of situation.


\begin{result}\label{Fine structure law}
	The approximate \emph{fine structure law} with $K$ additional neutrons for \molecule\, is equal to 
	{\small
		\begin{equation*}
			\mathrm{Multi} \left(
				\frac{ \lambda_\cem{^{13}C} }{ \mu }, 
				\frac{ \lambda_\cem{^{2} H} }{ \mu }, 
				\frac{ \lambda_\cem{^{15}N} }{ \mu },
				\frac{ \lambda_\cem{^{17}O} }{ \mu }, 
				\frac{ \lambda_\cem{^{33}S} }{ \mu }; 
				k_1
			\right ) \otimes
			\mathrm{Multi} \left(
				\frac{ \lambda_\cem{^{18}O} }{ \eta },
				\frac{ \lambda_\cem{^{34}S} }{ \eta }; 
				k_2	
			\right) \otimes 
			\mathbb{L}( k_1, k_2, k_4 ),
		\end{equation*}
	}
	where 
	\begin{equation}\label{simple lucky law}
		\mathbb{L}( k_1, k_2, k_4 ) = 
		\frac{ \frac{ \mu^{k_1} }{ k_1! } \frac{ \eta^{k_2}}{ k_2! } \frac{ \gamma^{k_4} }{ k_4! } }{ 
			\underset{ k_1' + 2 k_2' + 4 k_4' = K}{\sum} 
				\frac{ \mu^{k_1'} }{ k_1'! } 
				\frac{ \eta^{k_2'}}{ k_2'! } 
				\frac{ \gamma^{k_4'}}{ k_4'! }
		}.
	\end{equation}
\end{result}

Otherwise stated, the approximate distribution is a mixture of independent multinomial distributions weighted by the $\mathbb{L}$ distribution, which, for lack of name, we shall call the {\it lucky law}. Under the Poisson approximation, the {\it lucky law} is the resulting law on the {\it equatransneutronic configurations}. General expression is to be found in the \textbf{Appendix}.

Expression of type $\lambda_\cem{^{13}C}/\mu $ can have an interpretation of relative intensities of isotopes within a particular {\it equatransneutronic grouping}.

As pointed out in \cite{Olson2009Calculations}, it is of interest to calculate also the masses of the 
{\it equatransneutronic groups}. With Result \ref{Fine structure law}, we can provide extremely tractable approximations thereof. 
\begin{result}\label{Equatransneutronic Masses result}
	The approximate mass of a \emph{ transneutronic group} $(k_1, k_2, k_4)$ for compound \molecule\, is equal to
	{\small
		\begin{equation}\label{Equatransneutronic masses eq}
		\begin{gathered}
			\frac{k_1}{\mu}
			\Big( 
				{\Delta M}_\cem{^{13}C} \lambda_\cem{^{13}C} 	+ 
				{\Delta M}_\cem{^{2}H} 	\lambda_\cem{^{2}H} 	+ 
				{\Delta M}_\cem{^{15}N} \lambda_\cem{^{15}N} 	+ 
				{\Delta M}_\cem{^{18}O} \lambda_\cem{^{18}O} 	+  
				{\Delta M}_\cem{^{33}S} \lambda_\cem{^{33}S} 
			\Big)\\
			+\frac{k_2}{\eta}
			\Big( 
				{\Delta M}_\cem{^{18}O} \lambda_\cem{^{18}O} 	+ 
				{\Delta M}_\cem{^{34}S} \lambda_\cem{^{34}S} 
			\Big) + 
			\frac{k_4}{\gamma} {\Delta M}_\cem{^{36}S} \lambda_\cem{^{36}S} + 
			\textsc{Mono}_{ \cem{c}, \cem{h}, \cem{n}, \cem{o}, \cem{s} }, 
		\end{gathered}		
		\end{equation}}
	where $\Delta M$ stands for mass difference between a given isotope and the lightest isotope for that element, and \textsc{Mono} is the compound's monoisotopic mass.   
\end{result}
\begin{proof}
	It follows from the expression for multinomial law's mean, see \cite{Roos1999OnTheRateOfMultivariatePoissonConvergence}.
\end{proof}

Finally, note that other moments of the {\it equatransneutronic groupings} are readily obtained with the use of the multinomial moment generating function. 
\section{Algorithms}

Results \ref{Fine structure law} and \ref{Equatransneutronic Masses result} open up a new way to do calculations: using the approximation one reduces the complexity of Problem \ref{Problem of finding LFS_K configurations.} to that of studying $\mathbb{L}$. The {\it lucky law} is usually defined on a less dimensional space than $\MK$ and that significantly reduces the  computational effort. In proteomics, the state space of $\mathbb{L}$ can be thought to be two-dimensional, making it possible to establish the mass and probability of every {\it equatransneutronic grouping} in a {\it double for loop}. This approach is described as Algorithm \ref{DeFiner code}, code-named \textsc{DeFiner}. In general, exploration of $\mathbb{L}$ could be achieved by a tailored MCMC algorithm.   

\begin{algorithm}\caption{\textsc{DeFiner}}\label{DeFiner code}
\begin{algorithmic}\label{DeFiner Parallely}
	\State	\textbf{Input:}\,\,\,\,\, \molecule, $K$
	\State 	\textbf{Output:} A triplet of arrays with configurations, their probability, and mass.
	\State 	Establish $\lambda_\cem{^{13}C}, \lambda_\cem{^{2} H}, \lambda_\cem{^{15}N}, \lambda_\cem{^{17}O}, \lambda_\cem{^{33}S}, \mu, \gamma $, and all mass differences $\Delta M$.
	\State 	Find   $S = \{ (k_1, k_2, k_4) : k_1 + 2 k_2 + 4 k_4 = K \}$.
	\FORALL{ $\bm{k} \in S$}
		\State 	$\mathrm{Lucky}(\bm{k}) := \mathbb{L}\Big( \{ \bm{k} \} \Big)$
		\State 	$M(\bm{k}) :=$ mass of configuration $\bm{k}$ obtained using Eq. \eqref{Equatransneutronic masses eq} 
	\ENDFORALL
	\State 	Return $\{ S, \mathrm{Lucky}, M\}$.
\end{algorithmic}	
\end{algorithm}

\textsc{DeFiner} can be used as a subroutine for \textsc{DeFinest}: an algorithm that provides the exact multinomial peaks. \textsc{DeFinest} works as follows. 

First, having obtained the {\it lucky} configurations, we order them in descending $\mathbb{L}$-probability and select the critical $L\%$-set $S_L$ to trim out the asymptotically negligible configurations. To show it is so, let us introduce some extra notation
\begin{equation}
	\bm{g}_1 := (\cem{c_1}, \cem{h_1}, \cem{n_1}, \cem{o_1}, \cem{s_1} ), \quad
	\bm{g}_2 := (\cem{o_2}, \cem{s_2}), \quad
	g_4 := \cem{s_4}, \quad 
	\bm{g} := ( \bm{g}_1, \bm{g}_2, g_4)
\end{equation}
We think of $\bm{g}_1$ and $\bm{g}_2$ as of configurations of the multinomial laws described in Result \ref{Fine structure law}. Observe that entries of $\bm{g}_i$ sum to $k_i$. By Result \ref{Fine structure law}, note that $\QK \big( S_L \big) \leq \QK \big( S \big) = \mathbb{L}\big( \{ \bm{k} \} \big)$, where $\bm{k} := (k_1, k_2, k_4)$, the probability of a peak in a given {\it equatransneutronic grouping} being smaller than the probability of all the peaks gathered in it. Therefore, asymptotically all the $\bm{g}$ configurations in $S_L$ have a small $\QK$-probability, and we can decide whether to neglect them using only the information contained in $\mathbb{L}\big( \{ \bm{k} \} \big)$.

Subsequently, for each configuration $\bm{k}$ in $S_L$, one independently identifies critical $M\%$-set $\mathfrak{M}$ and critical $B\%$-set $\mathfrak{B}$ of the two underlying multinomial distributions. This can be achieved in many ways, see \textbf{Appendix} for our approach. With these sets at hand we calculate their exterior product and obtain a set of valid configurations from $LFS_K$. We then find their true $\MK$-probability and their mass $M$ using Eq. \eqref{Equatransneutronic masses eq}. We make use of \textsc{BRAIN} \cite{Dittwald2013BRAIN} software to get $\MM(LFS_K)$ needed to calculate $\MK$. Finally, we merge all obtained solutions.

Say that the algorithm resulted in set $A$ of configurations. One can measure \textsc{DeFinest}'s performance simply by calculating the overall $\textsc{Coverage} := \MK(A)$; the higher it is, the better we are in solving Problem \ref{Problem of finding LFS_K configurations.}.

\begin{figure}[htbp]
 \centering
 \includegraphics[width=.8\textwidth]{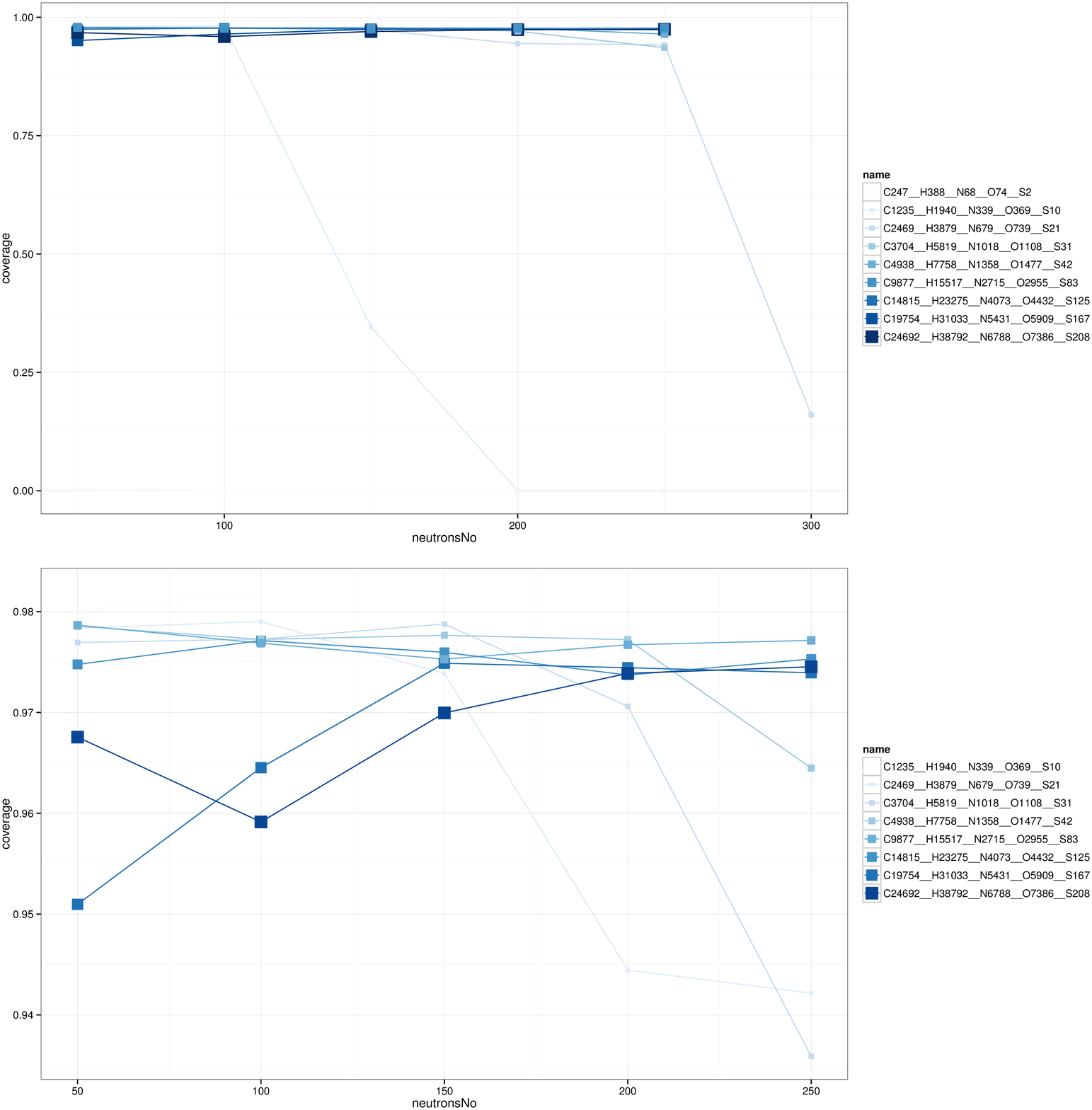}
 \caption{Coverage obtained using {\sc DeFinest} algorithm. The image on the bottom zooms into the upper reaches of the top picture. Both show the coverage of distribution original distribution $\MK$ for $K \in \{50, 100, 150, 200, 250, 300 \}$ for several chemical compounds. The bigger the compound (empirical formulas in the legend) the bigger the squares and the more intense the colour. Observe that for lighter compounds the results do not seem promising: we attribute this to the overall quality of conditional distributions $\MK$. Simply, all the multinomial distribution in \eqref{product of multinomials} are unimodal and for larger $K$ the solutions to Diophantine equation \eqref{LFS_K} do not encompass the region next to the mode, where the distribution is centered. For the reasons exposed in \textbf{Conclusions}, it is impractical to look at these distribution in the first place.}
 \label{figure: Coverage}
\end{figure}

A prototype of \textsc{DeFinest} has been implemented in \textbf{R}. Its pseudo code is described as Algorithm \ref{DeFinest code}. Observe also, that the {\it for loop} can be carried out in parallel. Fig. \ref{figure: Coverage} shows how well the prototype manages in solving Problem \ref{Problem of finding LFS_K configurations.}.

\begin{algorithm}\caption{\textsc{DeFinest}}\label{DeFinest code}
\begin{algorithmic}
	\State	\textbf{Input:} \molecule, $K, L, M, B$
	\State 	\textbf{Output:} array of masses and probabilities, \textsc{Coverage}.
	\State 	Run \textsc{DeFiner} and obtain $\{ S, \mathrm{Lucky}, M\}$.
	\State 	$S_L$:= top $L\%$ of configurations from $S$ ordered by their {\it lucky probabilities}, $\mathbb{L}(\{ \bm{k} \})$.

	\FORALL{ $\bm{k} \in S_L$}
		\State $\mathfrak{M}$ := Critical $M\%$ set of $\mathrm{Multi} \left(
				\frac{ \lambda_\cem{^{13}C} }{ \mu }, 
				\frac{ \lambda_\cem{^{2} H} }{ \mu }, 
				\frac{ \lambda_\cem{^{15}N} }{ \mu },
				\frac{ \lambda_\cem{^{17}O} }{ \mu }, 
				\frac{ \lambda_\cem{^{33}S} }{ \mu }; 
				k_1
			\right )$

		\State $\mathfrak{B}$ \,:= Critical $B\%$ set of 
		$\mathrm{Multi} \left(
			\frac{ \lambda_\cem{^{18}O} }{ \eta },
			\frac{ \lambda_\cem{^{34}S} }{ \eta }; 
			k_2	
		\right)$		

		\State $\mathfrak{R}_{\bm{k}}$ := $ 
			\Bigg\{ 
				\Big( 
					\MK( \bm{g} ), 
					M( \bm{g} )
				\Big) : \bm{g} = ( \bm{g}_1, \bm{g}_2, g_4 ) \in \mathfrak{M} \otimes \mathfrak{B} \otimes \{ k_4 \}  
			\Bigg\}$, cf. Eq. \eqref{Equatransneutronic masses eq} 
	\ENDFORALL

	\State 	Find {\sc Coverage}.
	\State 	Return $\big\{ \bigcup _{\bm{k}} \mathfrak{R}_{\bm{k}}, \textsc{Coverage} \big\}$.
\end{algorithmic}	
\end{algorithm}
\section{Conclusions}

	In the present paper an original approach to doing calculations on different levels of isotopic fine structure aggregation hierarchy was proposed. To our best knowledge, it is the first use of Poisson approximation for algorithmic purposes,   resulting already in two elegant algorithms, \textsc{DeFiner} and \textsc{DeFinest}, for efficient exploration of the state space of possible isotopic configurations.  

	\textsc{DeFiner} presents a minimalist, yet extremely efficient way to calculate the approximate probabilities of {\it equatransneutronic} clusters. \textsc{DeFinest} presents a simple, yet certainly suboptimal way of handing Problem \ref{Problem of finding LFS_K configurations.}; however, more efficient algorithms can easily come into being by more careful considerations on the structure of approximate distribution $\QK$. 

\begin{figure}[htbp]
 \centering
 \includegraphics[width=\textwidth]{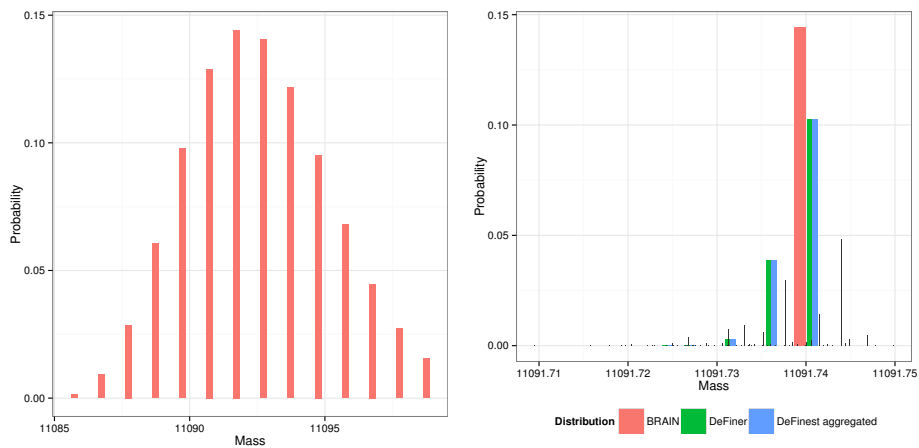}
 \caption{ Peaks in the left pane are probabilities of different $LFS_K$ groups, $K = 0,\dots,13$. In the right pane masses of configurations in $LFS_6$ are plotted: it zooms the region around the tallest peak in the left pane, which is also plotted there for reference. By appropriately aggregating \textsc{DeFinest}'s results, i.e. small black peaks, we calculate the {\it equatransneutronic} precise, non-approximated probabilities, in blue. We compare them with \textsc{DeFiner}'s results obtained {\it via} the Poisson approximation, in green. There are no apparent differences between them. }
 \label{figure: hierarchy}
\end{figure}

	Figure \ref{figure: hierarchy} presents a detailed view of the hierarchical approach we take. The left pane contains the  aggregated isotopic distribution of \testAvergine, an $100$-avergine, obtained with the {\sc BRAIN} algorithm \cite{Dittwald2013BRAIN}. The lower panel zooms into the region of the highest aggregated peak. This peak is then disaggregated into {\it equatransneutronic} groupings. Finally, one notices many small black peaks corresponding to the finest structure obtainable. It is by clustering and statistical centroiding of these peaks that one obtains all the others. 

	The potential applications of our results are numerous. Above all, the fine structure models can find application in automatic top-down peptide identification procedures by establishing more detailed fingerprints thereof and possibly boosting the ability to differentiate between similar compounds. Differences in the fine structure with $K^*$ s.t. $\mathbb{M}_{K^*}(LSF_{K^*}) = \max_K \MK(LSF_K)$ could be particularly informative.

	As another application, one can ask how to set up an optimal binning procedure. Simply, with a critical set of configurations $A$, s.t. $\MK(A) \approx 95\%$, how should these configurations be glued together to match real data from a mass spectrometer. This way, one could measure the machine's resolution without a need to refer to somewhat underdefined notions of {\it p percent valley} and {\it peak width}, see \cite{Eidhammer2008ComputationalMethodsInMassSpectrometry}.

	Observe that in this article we do not comment on the quality of the approximations in use. The reason behind it is that to our best knowledge no-one has ever carried out a thorough statistical research comparing which of these distributions is better suited for modelling the actual data. From the theoretical perspective, it seems plausible to adopt the most simple model of the isotopic fine structure probability, $\MM$, as developed in \cite{Kienitz1961MassSpectrometry}. However, with $\QQ$ at hand, and many data sets at disposal, one could verify whether such hypothesis holds. To our best knowledge, up to this moment only comparisons between theoretical distributions were carried out \cite{Valkenborg2007UsingPoisson}. We are of opinion that only through comparisons explicitly based on empirical data should one decide on the quality of the two models.

\subsubsection*{Acknowledgments.}

We would like to thank Alan Rockwood, Dirk Valkenborg, and, above all -- Piotr Dittwald, for fruitful discussions on the isotopic fine structure related issues.
  \bibliographystyle{splncs03.bst}
  \bibliography{spectrometry,probability,topology,generalMath,computerSciences}

\section*{Tables}

\begin{table}[ht]
	\centering
	\caption{Basic Information on Stable Isotopes, as found in \cite{Rosman1997IsotopicCompositions}.}\label{basic info on isotopes table}
	\begin{tabular}{lccll}
		\toprule
Element 	& Isotope 		& Extra Neutrons& Mass [Da] & Probability 	\\
		\midrule
\multirow{2}{*}{Carbon}  	
			& \ce{^{12}C} 	& 0 			& 12 		& 0.9893 		\\	
  			& \ce{^{13}C} 	& 1 			& 13.0033 	& 0.0107 		\\	
  		\cmidrule(r){1-5}
\multirow{2}{*}{Hydrogen}  	
			& \ce{^1H} 		& 0 			& 1.0078 	& 0.999885 		\\	
  			& \ce{^2H} 		& 1 			& 2.0141 	& 0.000115 		\\	
  		\cmidrule(r){1-5}	
\multirow{2}{*}{Nitrogen}  	
			& \ce{^{14}N} 	& 0 			& 14.0031 	& 0.99632 		\\	
  			& \ce{^{15}N}	& 1 			& 15.0001 	& 0.00368 		\\	  	  
  		\cmidrule(r){1-5}	
\multirow{3}{*}{Oxygen}  	
			& \ce{^{16}O} 	& 0 			& 15.9949 	& 0.99757 		\\	
  			& \ce{^{17}O}	& 1 			& 16.9991 	& 0.00038 		\\	  	  	
  			& \ce{^{18}O}	& 2 			& 17.9992 	& 0.00205 		\\	  	  
  		\cmidrule(r){1-5}	
\multirow{4}{*}{Sulfur}  	
			& \ce{^{32}S} 	& 0 			& 31.9721 	& 0.9493 		\\	
  			& \ce{^{33}S}	& 1 			& 32.9714 	& 0.0076 		\\	  	  
  			& \ce{^{34}S}	& 2 			& 33.9679 	& 0.0429 		\\
  			& \ce{^{36}S}	& 4 			& 35.9671 	& 0.0002 		\\		
		\bottomrule
	\end{tabular}
\end{table}

\section*{Figures}

\section*{Appendix}

\subsection*{Proof of Lemma \ref{conditional convergence lemma}}

We want to prove that if $\mu^{[n]} \rightharpoonup \mu$ and $\mu^{[n]}(A), \mu(A) > 0$, then also $\mu^{[n]}_A \rightharpoonup \mu_A$. We do this under the assumption that both $\mu^{[n]}$ and $\mu$ are discrete measures on probability space $E$. 

By the {\it Portmanteau Lemma}, see \cite{Kallenberg2002FoundationsOfModernProbability}, $\mu^{[n]} \rightharpoonup \mu$ implies that for any set $A$ with boundary $\partial A$ subject to $\mu( \partial A) = 0$, one should observe 

\begin{equation}\label{convergence in probability on good sets}
	\lim_{n \to \infty} \mu^{[n]}(A) = \mu(A).
\end{equation}

The notion of boundary requires the notion of topology: thus, we decide on the discrete topology, which is natural in this context \footnote{For appropriate topological notions consult \cite{Dugundji1966Topology}.}. In this topology however, $\partial A = \emptyset$, for it is a set theoretical difference of the closure and the interior, both of which are equal to $A$. Hence, $\mu( \partial A) = 0$. Thus, \eqref{convergence in probability on good sets} always holds.

{\it Ex definitione}, $\mu^{[n]} \rightharpoonup \mu$ means, that for any bounded function $f:E\to\mathbb{R}$ one observes
\begin{equation}\label{weak convergence definition}
	\int f \mathrm{d} \mu^{[n]} \underset{n \to \infty}{\xrightarrow{\hspace*{1cm}}} \int f \mathrm{d}\mu\,.
\end{equation}

A simple calculation using both \eqref{convergence in probability on good sets} and \eqref{weak convergence definition} completes the proof:

\begin{equation*}
	\int f \mathrm{d} \mu^{[n]}_A =  \frac{ \int f \mathrm{d} \mu^{[n]} }{ \mu^{[n]}(A) } \underset{n \to \infty}{\xrightarrow{\hspace*{1cm}}} \frac{ \int f \mathrm{d} \mu }{ \mu(A) } = \int f \mathrm{d} \mu\,.
\end{equation*}

\subsection*{General form of the \emph{Lucky Law}}

For \molecule, the parameters of the Diophantine equation defining $LFS_K$, see Eq. \eqref{LFS_K}, take values in set $I = \{ 1, 2, 4\}$. For a general set $\mathcal{I}$, formula \eqref{simple lucky law} generalizes to 
\begin{equation*}
	\mathbb{L}( \bm{k} ) = 
	\frac{ 
		\prod_{i \in \mathcal{I}} \frac{ \mu_i^{k_i} }{ {k_i}! } 
	}{ 
		\underset{ \{ \bm{k}^* :  \sum_{i \in \mathcal{I}} i k_i^*  = K \} }{\sum} 	
		\prod_{i \in \mathcal{I}} \frac{ \mu_i^{k_i^*} }{ {k_i^*}! }	
	},
\end{equation*}
where $\bm{k}$ is an ordered tuple indexed by $\mathcal{I}$. Nature poses a natural limit on the complexity of the {\it lucky law}, as at most  $\# \mathcal{I} \leq 10$. Observe also, that this law arises from conditioning a product of independent $\#\mathcal{I}$ Poisson distributions conditioned on the Diophantine equation $\sum_{i \in \mathcal{I}} i k_i^*  = K$.

\subsection*{Obtaining $M\%$ critical sets of the Multinomial Distribution}

Stating the algorithm requires some extra notation: let $\mathfrak{S}_k = \{ \bm{c} = (c_1, \dots, c_w) : \sum_{i=1}^w c_i = k, c_i \geq 0 \}$, a simple $k-$simplex, be the underlying state-space for the multinomial distribution, $\mathcal{M} := \mathrm{Multi}( p_1, \dots, p_w ; k)$. We can then consider a graph $G = (V,E)$, where the set of vertices $V \equiv \mathfrak{S}_k$ and with edges $E$ specified as follows: two configurations $\bm{a}, \bm{b} \in V$ form an edge $(\bm{a}, \bm{b}) \in E$ if and only if $\exists_{i,j \in \{ 1, \dots, w \}, i \not= j} a_i = b_i + 1$ and $a_j = b_j - 1$. 

The algorithm amounts then to performing a controlled {\it breadth first search}. We start the search in the vicinity of $\mathcal{M}$'s mode, using as proxy point $\bm{c}$ with coordinates set as $ c_i := n p_i + 1$. More elaborate set of candidates can be used, see \cite{Gall2003DeterminationOfTheModesOfMultinomial}. We then enlists all $\bm{c}$'s neighbours and puts them altogether on a {\it max-priority queue}, see \cite{Cormen2001IntroductionToAlgorithms}. We then look at neighbours of the top-priority configuration, check their probability and enqueue them. In the meantime, we store information on the visited configurations in a hash table to avoid multiple visits to the same node. We collect information about the total probability of the already visited nodes and their number. We stop the algorithm as soon as the accumulated probability reaches a number greater than the prespecified threshold level $M$ or if the number of already observed peaks reaches a prespecified number, i.e. when there will be too many peaks.

Observe that in case of molecules containing elements with only one isotope, e.g. \smallMolecule, the above algorithm suffices to solve the Problem \ref{Problem of finding LFS_K configurations.}, as showed in Result \ref{Multinomial Result}.

\end{document}